\documentclass[12pt]{article}
\usepackage{a4wide}
\usepackage{epsfig}
\usepackage{amssymb}
\usepackage{graphicx}
\usepackage{amsmath}
\usepackage[center]{subfigure}

\newcommand{\ba}{\begin{eqnarray}}
\newcommand{\ea}{\end{eqnarray}}
\newcommand{\be}{\begin{equation}}
\newcommand{\ee}{\end{equation}}
\begin{document}
\begin{titlepage}
\begin{flushright}
SI-HEP-2005-01\\
\end{flushright}
\vfill
\begin{center}
{\Large\bf 
$B$-Meson Distribution Amplitude\\ 
from the $B\to \pi$ Form Factor}\\[2cm]
{\large\bf  
Alexander~Khodjamirian, Thomas Mannel and 
Nils Offen }\\[0.5cm]
{\it  Theoretische Physik 1, Fachbereich Physik,
Universit\"at Siegen,\\ D-57068 Siegen, Germany }\\
\end{center}
\vfill
\begin{abstract}
Employing the light-cone sum rule approach in QCD, we relate the $B$-meson distribution amplitude
to the $B\to \pi$ form factor at zero momentum transfer. 
In leading order, the sum rule is converted  into a simple 
expression for  the inverse moment $\lambda_B$  of 
the distribution amplitude $\phi_{+}^B$.  
Using as an input  the $B\to\pi$ form factor 
calculated from the light-cone sum rule in terms 
of pion distribution amplitudes, we obtain an estimate:
$\lambda_B =460 \pm 160 $ MeV. 
We investigate how this result is modified by the 
$B$-meson three-particle distribution amplitudes.

\end{abstract}
\vfill

\end{titlepage}

\newcommand{\DS}[1]{/\!\!\!#1}

\section{Introduction}

The $B$-meson distribution amplitudes (DA's)
emerge as universal nonperturbative objects 
in many studies  of exclusive $B$ meson decays 
\cite{BHS,NG,BF},
in particular, in $B\to\gamma l \nu_l$ \cite{blnugamma}. 
Still there is a very limited knowledge  of the 
nonperturbative parameters determining 
these DA's. Especially important is the inverse moment 
of the two-particle DA $\phi^B_+$ (see the definition below)
which enters many factorization formulas, including 
QCD factorization for hadronic  \cite{BBNS}
and radiative  \cite{BVgamma} $B$ decays.  

As usual, when there is a need to calculate some 
hadronic parameter,  the method of QCD sum rules \cite{SVZ}
turns out to be helpful. The ``classical'' 
two-point sum rule, based on the local OPE and 
condensate  expansion, was used for the $B$-meson DA's 
already in the original study \cite{NG}. Recently, the
two-point sum rule calculation was improved in \cite{BIK},
including next-to-leading order corrections.  An estimate 
of the inverse moment of $\phi^B_+$ was also obtained 
in \cite{BallKou}, matching the factorization 
formula to the light-cone sum rule for 
$B\to\gamma l \nu_l$. Currently, the spread 
of model assumptions and sum rule predictions 
for the $B$ meson DA's in the literature 
remains rather large. An additional QCD
estimate, even an approximate one, can be useful in that
respect.      

In this paper we suggest a new approach, using 
the light-cone sum rule technique \cite{lcsr} 
and relating the 
$B$-meson DA to the $B\to \pi$ form factor. The latter  
is a better known hadronic object,
receiving a lot of attention in recent years. 
At small momentum transfers this form factor  was  
calculated in QCD using light-cone sum rules \cite{BpiLCSR}, 
based on OPE in terms of the pion DA's with growing twist. 
Therefore, the link between the $B$-meson DA and 
$B\to \pi$ form factor established below,
provides an independent dynamical information 
on the $B$-meson DA.

The aim of this letter is mainly to outline the procedure 
and to demonstrate that it works at the leading-order level.
Gluon effects originating from the three-particle Fock states 
of $B$ meson are also investigated at a 
qualitative level, within our very limited knowledge  
of the quark-antiquark-gluon DA's. 
Perturbative corrections remain  beyond the 
scope of this work, being postponed to 
a future study. Hence we also do not take into account the 
nontrivial renormalization  of the 
$B$-meson DA worked out  in \cite{NL} (see also \cite{BIK}).

\section{ Correlation function}
  
The starting object in our approach is the correlation function 
which represents a product of the heavy-light and light 
quark currents  sandwiched between the vacuum and 
the $B$-meson states:
\begin{equation}
F_{\mu\nu}^{(B)}(p,q)= i\int d^4x ~e^{i p\cdot x}
\langle 0|T\left\{\bar{d}(x)\gamma_\mu \gamma_5 u(x), 
\bar{u}(0)\gamma_\nu b(0)\right\}|\bar{B}^0(p+q)\rangle\,.
\label{eq-corr}
\end{equation}
This expression resembles 
the correlation function used 
in deriving the light-cone sum rules for $B\to \pi$ 
form factor  \cite{BpiLCSR}, with the roles of $B$ and 
$\pi$ interchanged:
in (\ref{eq-corr}) the $B$ meson is on shell, $(p+q)^2=m_B^2$, 
and the pion is interpolated by the axial-vector current. 
If the momentum squared of the axial-vector current 
is spacelike and sufficiently large:
\be
p^2<0,~ ~ |p^2|\sim 1~ \mbox{GeV}^2 \gg \Lambda_{QCD}^2\,, 
\label{cond1}
\ee
the intermediate $u$ quark is highly virtual and propagates 
near the light-cone $x^2= 0$. The momentum transfer squared 
in the heavy-light vertex  will be put to zero ($q^2=0$). 
In this kinematical configuration,
the product of currents in the correlation function (\ref{eq-corr}) 
can be expanded near the light-cone. Note that,  
up to the differences in the quantum numbers of currents, 
the correlation function (\ref{eq-corr}) is similar to the 
$B\to \gamma l \nu_l$ amplitude. 
For the latter, the light-cone dominance is usually assumed
(in the framework of the heavy-quark mass expansion) 
\cite{blnugamma} at the small mass squared $q^2$ of the lepton pair, 
(that is, at large photon energy
in the $B$-meson rest frame). However, the long-distance effects 
for the real photon ($p^2=0$), although $1/m_b$ suppressed, 
are still present \footnote{In the LCSR approach to 
$B\to \gamma l \nu_l$ \cite{BallKou,Brhogamma} the long-distance photon 
emission is described by the photon DA and at finite $m_b$ 
turns out to be large.}. With our choice (\ref{cond1}) these effects
in the correlation function (\ref{eq-corr}) are additionally 
suppressed by inverse powers of $|p^2|$. 

In the leading order, 
only one tree-level diagram contributes  (Fig. 1a), 
where the virtual $u$ quark is replaced by the free propagator:
\begin{equation} 
\langle 0 \mid T \{ u_\alpha (x) \bar{u}_\beta(0)\}\mid 0 \rangle 
=\frac{i\DS x_{\alpha\beta}}{2\pi^2 (x^2)^2}\,.
\label{eq-prop}
\end{equation}
Contracting the $u$-quark fields in (\ref{eq-corr}), 
we encounter the vacuum-$B$-meson matrix element 
of a bilocal operator, 
with the light-antiquark and heavy-quark fields at a 
near-to-light-cone separation. 
This hadronic matrix element 
is decomposed in the $B$-meson two-particle DA's 
\cite{NG,BF} we are interested in : 
\begin{eqnarray}
&&
\langle 0|\bar{d}_\alpha(x)[x,0] b_\beta(0)
|\bar{B}^0(v)\rangle
\nonumber \\ 
&&
= -\frac{if_B m_B}{4}\int\limits _0^\infty 
d\omega e^{-i\omega v\cdot x} 
\left [(1 +\DS v)
\left \{ \phi^B_+(\omega) -
\frac{\phi_+^B(\omega) -\phi_-^B(\omega)}{2 v\cdot x}\DS x \right \}\gamma_5\right]_{\beta\alpha} \,,
\label{eq-BDAdef}
\end{eqnarray}
where $v$ is the $B$-meson velocity ($p+q=m_Bv$), ~$[x,0]$ is the
path-ordered gauge factor,  
and $\phi_{\pm}^B$ are the distribution amplitudes 
in the momentum space.
The variable $\omega>0$ is the plus component of the 
momentum of the light spectator quark in the $B$ meson. Note that 
the DA's are essentially concentrated around
$\omega\sim \bar{\Lambda}$, where $\bar{\Lambda}=m_B-m_b$.

Neglecting the three-particle Fock states 
of $B$ meson, one obtains 
\cite{BF,KKQT} a simple differential relation 
of the Wandzura-Wilczek (WW) type between the 
two functions $\phi_{+}^B$ and $\phi_{-}^B$, which 
can also be written in the following form:
\be 
\phi_{-}^B(\omega)=\int\limits_\omega^\infty \!\!d\rho 
\,\,\frac{\phi_{+}^B(\rho)}{\rho}\,.
\label{eq-WW}
\ee

Using the definition  (\ref{eq-BDAdef}), we easily derive the 
leading-order answer for the light-cone expansion of 
the correlation function (\ref{eq-corr}). In what follows, we are 
interested in the invariant amplitude multiplying $p_\mu p_\nu$.  
After rearranging the part of the DA containing
$v\cdot x$ in the denominator by means of partial integration,
the answer for this amplitude turns out
to be surprisingly simple, in particular, the parts proportional
to $\phi_+^B$ cancel each other. The result reads:
\begin{equation}
F^{(B)}_{\mu \nu}(p,q)= 2if_B \int\limits_0^\infty 
\frac{d\omega}{m_B\omega-p^2} \phi_-^B(\omega)\,p_\mu p_\nu + ...\,,
\label{eq-LO_FB}
\end{equation}
where ellipses denote the remaining Lorentz structures.

\begin{figure}
\centering
\vspace{-1cm}
\includegraphics[width=0.7\textwidth]{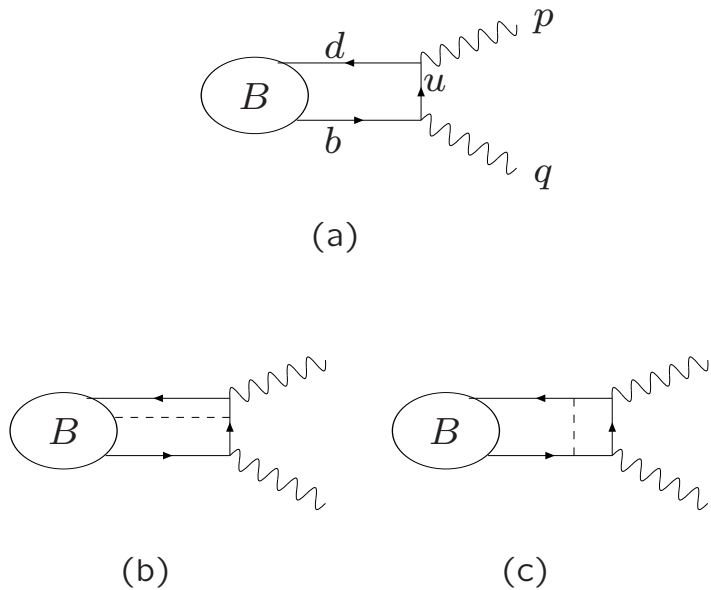}
\caption{
\it Diagrams 
corresponding to the correlation function (\ref{eq-corr}): 
(a) the leading-order diagram, (b) the soft-gluon emission 
with 3-particle DA's, (c) one of the diagrams with 
the hard gluon exchange. The wavy (dashed) lines
denote external currents (gluons).
}
\end{figure}

\section{Sum rule for the inverse moment}

The OPE  result (\ref{eq-LO_FB}) for the correlation function (\ref{eq-corr})  
will now be used to derive LCSR for $B\to \pi$ form factor.
The procedure is similar  to  the treatment of 
the vacuum-pion correlation function in \cite{BpiLCSR}.
The roles of the pion and $B$ meson are now reversed: 
the pion is generated by the interpolating current and has to be 
approximated
by quark-hadron duality, whereas the on-shell $B$ meson is 
represented by its DA.

We write down a hadronic dispersion relation in the channel 
of the axial-vector current with the momentum 
squared $p^2$.
The set of hadronic states contributing to that relation 
includes the pion, $a_1$ meson, excited
resonances and continuum states with $J^{PC}=0^{-+},
1^{++}$. 
Isolating the ground-state pion contribution from the rest, 
we obtain, at $q^2=0$:
\ba 
F^{(B)}_{\mu\nu}(p,q)= 
\left \{\frac{2if_\pi f^+_{B\pi}(0)}{-p^2}+ 
\int\limits_{s_h}^\infty ds \frac{\rho ^h(s)}{s-p^2}\right\}p_\mu p_\nu+... \,,
\label{eq-hadrdisp}
\ea
where the standard definitions of the pion decay constant 
$\langle 0|\bar{d}\gamma_\mu \gamma_5 u  |\pi(p)\rangle =
ip_\mu f_\pi$ and 
$B \to \pi$ form factors 
\be
\langle \pi(p)|\bar{u}\gamma_\nu b |B(p+q)\rangle =
(2p+q)_\nu f^+_{B\pi}(q^2) +q_\nu f^-_{B\pi}(q^2)
\label{Bpiformf}
\ee
are used.
Furthermore, in (\ref{eq-hadrdisp}) we neglect the pion mass and denote by $\rho^h$ the 
spectral density of all states heavier than the pion, 
with the corresponding hadronic threshold $s_h$.
In the channel of the axial-vector current 
quark-hadron duality works reasonably well, 
as known already from 
the original paper \cite{SVZ} on QCD sum rules.
In particular,  $f_\pi$ is reproduced
if one attributes to the pion  
the integral over the spectral density 
calculated from OPE with the threshold  
$s_0^\pi =0.7$ GeV$^2$ (see also the review \cite{CK}). 
The same duality ansatz was used in LCSR for the pion 
e.m. form factor \cite{pionff,BK}.

Note that  the expression (\ref{eq-LO_FB}) for the amplitude 
$F^{B}_{\mu\nu}$ is easily transformed into a form 
of dispersion relation,
if one replaces the variable $\omega$ by $s=m_B\omega$:
\be
F^{(B)}_{\mu \nu}= \left \{\int\limits_0^\infty 
\frac{ds}{s-p^2} \rho^{OPE}(s)\right\}p_\mu p_\nu + ...,~~ 
\mbox{with}~~\rho^{OPE}(s)= \frac{2if_B}{m_B}\phi_-^B(s/m_B)\,.
\label{qcddisp}
\ee
Equating  the above result to 
the hadronic dispersion relation (\ref{eq-hadrdisp})
at large spacelike $p^2$, we employ the quark-hadron 
duality approximation:
$\rho^h(s)\Theta(s-s_h)=\rho^{OPE}(s)\Theta(s-s_0^\pi)$ 
and apply the Borel transformation 
to both parts of this equation. This transformation
replaces $|p^2|$ with the effective scale $M^2\sim 1$ GeV$^2$. 
Simultaneously, the Borel exponent
suppresses the contribution of heavier than pion 
hadronic states in the dispersion integral, so that  the 
resulting sum rule is less sensitive to the accuracy 
of the duality approximation. Finally, we obtain:
\be    
\int\limits_0 ^{s_0^\pi} ds e^{-s/M^2} 
\phi_{-}(s/m_B) \simeq  \phi_{-}(0)\int\limits _0 ^{s_0^\pi} 
ds e^{-s/M^2}=\frac{f_\pi f^+_{B\pi}(0)}{f_B}m_B\,,
\label{eq-SR}
\ee
where we used the fact that $\sqrt{s_0^\pi}\sim \bar{\Lambda}$ 
and $s^0_\pi \ll m_B^2$.

This relation has a very transparent physical 
interpretation: in $B\to \pi$ transition 
the pion (in the absence of hard gluon exchange) is produced 
via end-point mechanism, 
that is, picking up the light spectator-quark of $B$ meson 
with a vanishing  momentum fraction.    

Using the relation between $\phi_{-}(0)$ and $\phi_{+}$
in the form (\ref{eq-WW}) we encounter on l.h.s. 
of (\ref{eq-SR}) the inverse
moment defined as 
\be
\int\limits _0 ^\infty d\omega \frac{\phi_{+}(\omega)}{\omega}=\frac{1}{\lambda_B}\,.
\label{eq-inverse}
\ee
Hence, an approximate relation for $\lambda_B$ can 
be established:
\be
\frac{1}{\lambda_B}= \frac{f_\pi f^+_{B\pi}(0)m_B}{f_B M^2(1-
e^{-s_0^\pi/M^2})}\,,
\label{eq-rel}
\ee
which is valid up to unaccounted $O(\bar{\Lambda}/m_B)\sim O(\sqrt{s_0^\pi}/m_B)$ corrections.
The normalization scale $\mu$ of $\lambda_B$  
in this relation can only be roughly estimated.
Since the average 
virtuality of the intermediate $u$-quark in the correlator 
is between $M^2$ and  $\bar{\Lambda} m_b$, 
we expect that 1 GeV $<\mu <\sqrt{\bar{\Lambda} m_B}$.

\section{Numerical estimate of $\lambda_B$}

To obtain a numerical estimate of $\lambda_B$ from (\ref{eq-rel}), 
we take as an input  the threshold parameter  
$s_0^\pi =0.7$ GeV$^2 $ and the interval 
$M^2=0.5 \div 1.2$ GeV$^2$, used in 
the QCD sum rule for the pion channel \cite{SVZ,CK} .

Concerning $B\to\pi$ form factor and $f_B$, it is natural
to take the estimates obtained from LCSR (in terms of pion DA's) 
and from the two-point QCD sum rule, respectively. 
Strictly speaking, since the radiative corrections are not yet
included, the quantity $f^+_{B\pi}(0)$  in (\ref{eq-rel}) 
has to be interpreted as the ``soft'' (end-point or nonfactorizable)
part of the form factor. On the other hand, the LCSR analysis
\cite{BpiLCSR} predicts that at finite $m_b$ 
the ``hard-scattering''(factorizable) part of the $B\to \pi$ 
form factor determined by the radiative corrections to the sum rule, 
is subdominant (see also \cite{BallSCET}). Therefore, we  
take the complete form factor predicted from LCSR.
More specifically, to obtain the ratio 
$f^+_{B\pi}(0)/f_B$ entering (\ref{eq-rel}),
we calculate the product  $f^+_{B\pi}(0)f_B$
from LCSR and divide it by the two-point sum rule 
for $f_B^2$, both taken with $O(\alpha_s)$ accuracy. 
This calculation uses 
the same input as in \cite{KRWWY} (see also \cite{KMM}) 
and, in particular, yields $f^+_{B\pi}(0)=0.26 \pm 0.06 $ 
\footnote{ This interval agrees with the recent 
LCSR estimate $f^+_{B\pi}(0)=0.258 \pm 0.031$ \cite{BallZw}.}.
The remaining parameters in (\ref{eq-rel}) are  
$m_B=5.279$ GeV  and $f_\pi=131 $ MeV.     
Our final result is 
\be
\lambda_B  =460   \pm  160~\mbox{MeV}\,, 
\label{eq-lambN}
\ee
where the uncertainties due to the variation 
of input parameters are added in quadrature.
Including gluonic corrections to 
(\ref{eq-rel}) will allow one to improve 
the above estimate. However the achievable 
accuracy of  the $B\to \pi$ form factor and $f_B$ puts a lower 
limit of about $\pm$ 20\% on the uncertainty in this relation.
 
Our prediction reveals an encouraging agreement with  
the result of the 2-point sum rule calculation 
\cite{BIK} $\lambda_B =460   \pm  110~\mbox{MeV}$,
being also consistent with 
$\lambda_B= 350 \pm 150$ MeV
adopted in QCD factorization approach \cite{BBNS} 
and with the  estimate $\lambda_B\simeq 600$ MeV  
inferred \cite{BallKou} from a LCSR for $B\to \gamma l \nu_l$.

\section{Gluon Corrections}

The three-particle (quark-antiquark-gluon) Fock states of 
$B$ meson  influence the relation (\ref{eq-rel})
in two ways: 1) directly, via the gluon emission from the virtual 
quark line (diagram in Fig.~1b), 
and 2) indirectly, due to modification of the WW relation
(\ref{eq-WW})
. 
Since very little is known about the $B$-meson three-particle DA's, 
our analysis of both  effects presented in this section is rather 
qualitative.

We calculated the diagram in Fig.~1b, inserting in the 
correlator (\ref{eq-corr}) the one-gluon part of the
$u$-quark propagator near the light-cone \cite{BB}. 
Instead of Eq.~(\ref{eq-prop}) one then uses
\begin{equation} 
\langle 0 \mid T \{ u_\alpha (x) \bar{u}_\beta(0)\}\mid 0 \rangle 
=-\frac{i}{16\pi^2x^2}\int\limits_0^1 du \,G_{\tau\rho}(ux)\left(
\DS x \sigma^{\tau\rho}-4iux_\tau\gamma_\rho\right)_{\alpha \beta}\,,
\label{eq-prop1}
\end{equation}
where $G_{\tau\rho}=g_sG_{\tau\rho}^a(\lambda^a/2)$
and the Fock-Schwinger
gauge is adopted, with the path-ordered gauge factor equal to the unit.
The vacuum-to-$B$ matrix element is then decomposed 
in four independent  three-particle DA's 
introduced in  \cite{KKQT} in terms of the most general 
parameterization compatible with the heavy-quark limit:
\begin{eqnarray}
&&\langle 0|\bar{d}_\alpha(x) G_{\lambda\rho}(ux) 
b_\beta(0)|\bar{B}^0(v)\rangle=
\frac{f_Bm_B}{4}\int\limits_0^\infty d\omega
\int\limits_0^\infty d\xi\,  e^{-i(\omega+u\xi) v\cdot x} 
\nonumber \\ 
&&\times \Bigg [(1 +\DS v) \Bigg \{ (v_\lambda\gamma_\rho-v_\rho\gamma_\lambda)
\Big(\Psi_A(\omega,\xi)-\Psi_V(\omega,\xi)\Big)
-i\sigma_{\lambda\rho}\Psi_V(\omega,\xi)
\nonumber\\
&&-\left(\frac{x_\lambda v_\rho-x_\rho v_\lambda}{v\cdot x}\right)X_A(\omega,\xi)
+\left(\frac{x_\lambda \gamma_\rho-x_\rho \gamma_\lambda}{v\cdot x}\right)Y_A(\omega,\xi)\Bigg\}\Bigg]_{\beta\alpha}\,.
\label{eq-B3DAdef}
\end{eqnarray}
In the above, $\Psi_{V}$,$\Psi_{A}$, $X_A$ and $Y_A$ are the DA's in the momentuum space, the variables $\omega>0$ and $\xi>0$ being, respectively, the plus components of the light-quark and gluon momenta 
in the $B$ meson. 

In the following, we only take into account the first two 
DA's, $\Psi_V$ and $\Psi_A$. The contribution of the remaining 
two DA's to the sum rule is suppressed, at least by the inverse power 
of the Borel parameter. The result of our calculation
yields the following addition to the correlation function 
(\ref{eq-LO_FB}):
\ba 
F^{(B)\{\Psi_{V,A}\}}_{\mu \nu}(p,q)= 2if_B 
\int\limits_0^\infty d\omega\int\limits_0^\infty d\xi
\int\limits_0^1 du \frac{m_B}{(m_B-\omega-u\xi)(m_B(\omega+u\xi)-p^2)^2}\nonumber\\ 
\Big(\Psi_V(\omega,\xi)+(1-2u)\Psi_A(\omega,\xi)\Big)p_\mu p_\nu + ...\,,
\ea
\label{eq-corr3}
We assume that both $\Psi_V$ and $\Psi_A$ vanish at large 
$\omega,\xi$, so that 
all integrals are ultraviolet convergent \footnote{Since we work 
in LO approximation, the effect of ``radiative tail''\cite{NG,NL,BIK} 
originating from the hard gluon emission is neglected here.}.  
Furthermore, employing partial integration, we transform the 
above expression 
to the dispersion form and apply duality approximation as well as 
Borel transformation. The sum rule (\ref{eq-SR}) becomes:
\ba
\int\limits_0 ^{s_0^\pi} ds\, e^{-s/M^2} 
\Bigg\{ \phi_{-}(s/m_B) +
\frac{d}{ds}\Bigg[\frac{m_B^2}{m_B^2-s}\int\limits_0^{s/m_B}d\omega
\int\limits_{s/m_B-\omega}^{\infty}\frac{d\xi}{\xi}\Bigg(\Psi_V(\omega,\xi)
\nonumber\\
+
\left[ 1-2\left(\frac{s-m_B\omega}{\xi m_B}\right)\right]\Psi_A(\omega,\xi)\Bigg)\Bigg]\Bigg\}=\frac{f_\pi f^+_{B\pi}(0)}{f_B}m_B\,.
\label{eq-SR1}
\ea
To assess the relative role of the gluon correction, 
it is sufficient to establish the behavior of both DA's, 
$\Psi_V$ and $\Psi_A$,  
at small $\omega$ and $\xi$. We follow the same approach as 
in \cite{NG} and compare the decomposition (\ref{eq-B3DAdef}) 
valid in the heavy quark limit with the well known 
definitions of quark-antiquark-gluon DA's 
$\varphi_{3P}(\alpha_i)$,
$ \varphi_{\perp}(\alpha_i)$,
$\varphi_{\parallel}(\alpha_i)$,
$\widetilde{\varphi}_{\perp}(\alpha_i)$,
and $\widetilde{\varphi}_{\parallel}(\alpha_i)$,
of a pseudoscalar meson ${\cal P}$ with finite mass, where 
$\alpha_i\equiv \alpha_1,\alpha_2,\alpha_3$ and
$\alpha_1$,$\alpha_3$ are the fractions of the meson 
momentum carried by the light quark and gluon, respectively.
For example $\varphi_{3P}$ is defined as 
\ba
&&\langle 0|\bar{d}(x) G_{\lambda\rho}(ux)\sigma_{\mu\nu} 
b(0)|{\cal P}(P)\rangle
=if_{3P}\Big[\Big(P_\lambda P_\mu g_{\rho \nu}-
P_\lambda P_\nu g_{\rho \mu}\Big)
-\{\rho\leftrightarrow \lambda\}\Big]
\nonumber \\ &&\times
\int\limits_0^1 d\alpha_1 \!\int\limits_0^{1-\alpha_1}d\alpha_3\,
e^{-i(\alpha_1+u\alpha_3)P\cdot x}
\varphi_{3P}(\alpha_i)|_{\alpha_2=1-\alpha_1-\alpha_3}\,.
\label{phi3P}
\ea
where $f_{3P}$ is the nonperturbative normalization 
parameter, $P$ is the meson momentum and DA has the 
following asymptotic form:
\be
\varphi_{3P}(\alpha_i)\sim \alpha_1\alpha_2\alpha_3^2\,.
\label{asy}
\ee
For the sake of brevity we do not quote here definitions
and asymptotic forms of all other DA's, 
they can be found,  e.g. in the Appendix  B of \cite{BK}.
It is important that 3-particle DA's 
are proportional at least to the first power of the gluon 
momentum fraction $\alpha_3$.
From that we expect that, at $\xi \to 0$, also 
$\Psi_{A,V}(\xi)\sim \xi^ n$ with $n\geq 1$.
The latter condition is sufficient
for the infrared convergence of the 
integrals over $\omega,\xi$ in (\ref{eq-SR1}). Furthermore, we find 
that the gluon correction in (\ref{eq-SR1}) is  
at least $\sim 1/m_B$ suppressed with respect to the 
leading-order term proportional to $\phi_-$. 

Our final comment concerns the violation of the WW relation
which, in the presence of three-particle DA's, acquires
a correction worked out in \cite{KKQT}. In particular 
the relation for the inverse moment modifies to:
\be
\phi_{-}(0)=\frac{1}{\lambda_B} 
+\int\limits_0^\infty\frac{d\omega}\omega I(\omega)
\label{eq-invers1}
\ee
where 
\be
I(\omega)=2\frac{d}{d\omega}\int\limits_0^\omega d\rho
\int\limits_{\omega -\rho}^\infty \frac{d\xi}{\xi}
\frac{\partial}{\partial \xi}\Big(\Psi_V(\rho,\xi)-\Psi_A(\rho,\xi)\Big )
\ee
To deduce the low-energy behavior of the DA's  
in this integral one may again rely 
on comparison with the 3-particle DA's  at finite mass.
Multiplying (\ref{eq-B3DAdef})
by  $\sigma_{\mu\nu}\gamma_5$ 
and taking trace, we found that the resulting kinematical 
structure coincides with the one in (\ref{phi3P})
. Hence, we assume that the behavior at small
$\omega/m_B\to 0 $ and $\xi/m_B\to 0 $ of 
$(\Psi_V(\omega,\xi)-\Psi_A(\omega,\xi))$ is the same as the 
behavior of $\varphi_{3P}(\alpha_i)$ at $\alpha_1\to 0$ and 
$\alpha_3\to 0$, respectively, yielding, according to (\ref{asy}):
\ba
\Psi_V(\omega,\xi)-\Psi_A(\omega,\xi) \sim \varphi_{3\pi}
(\alpha_1,1-\alpha_1-\alpha_3, \alpha_3)|_{\alpha_1=\omega/m_B, ~\alpha_3=\xi/m_B}\sim \omega \xi^2\,,
\label{eq-comp}
\ea
so that  the integral in (\ref{eq-invers1})
is convergent. In order to estimate its numerical
value we use the normalization factors
\ba  
\int \limits_0^{\infty}d\omega\int\limits_0^{\infty} d\xi\,\,\Psi_V(\omega,\xi)=\frac{\lambda_H^2}{3}=(0.18\pm 0.07)/3
~\mbox{GeV}^2\,,
\nonumber\\
\int\limits_0^{\infty} d\omega\int\limits_0^{\infty} d\xi\,\, \Psi_A(\omega,\xi)=\frac{\lambda_E^2}{3}=(0.11\pm 0.06)/3
~\mbox{GeV}^2\,,
\label{lambdas}
\ea
estimated in \cite{NG} by relating them to the 
matrix elements of certain local operators
and calculating  these matrix elements from the sum rules  in HQET. 

We adopt a simple model for the difference of the two DA's:
\be
 \Psi_V(\omega,\xi)-\Psi_A(\omega,\xi)=
\frac{(\lambda_H^2-\lambda_E^2)}{6\bar{\Lambda}^5}\omega\xi^2
exp\left(-\frac{\omega+\xi}{\bar{\Lambda}}\right)\,, 
\label{model}
\ee
which has a typical ultraviolet behavior 
of the models for $\phi_{\pm}(\omega)$ adopted in \cite{NG}
and obeys both the low-energy limit (\ref{eq-comp})
and the normalization conditions (\ref{lambdas}).
Calculation of the gluon correction in 
(\ref{eq-invers1}) with (\ref{model}) then yields
(at $\bar{\Lambda}=0.6$ GeV):
\be
\int\limits_0^\infty\frac{d\omega}\omega I(\omega)=
\frac{\lambda_H^2-\lambda_E^2}{9\bar{\Lambda}^3}\simeq 
0.04 ~\mbox{GeV}^{-1}\,.
\ee
We see that this correction is only at the level of $2\%$ 
of our estimate for the leading order term 
in (\ref{eq-invers1}): $1/\lambda_B\sim2$ GeV$^{-1}$. 

Hence, soft gluon effects are most probably suppressed 
both in the sum rule and in the WW relation. 
A future, more refined analysis of the $B$-meson 3-particle DA's 
using dedicated QCD sum rules (similar to one used in \cite{BIK}
for $\psi_+$) will allow to estimate these effects
more accurately.

\section{Conclusion}

Summarizing, we  have suggested a new approach to the $B$-meson DA, by 
relating it to the combination of two hadronic observables: 
the $B\to \pi$ form factor at zero momentum transfer and $f_B$.  
The main result obtained in the zeroth order in $\alpha_s$ 
and neglecting soft gluon correction, is the sum rule 
for DA $\phi^B _{-}$, which is converted into an equation
for the inverse moment $\lambda_B$ of $\phi^B _{+}$ using the WW relation between these two DA's. Our numerical estimate of $\lambda_B$
agrees with the prediction of the two-point sum rule. 

In order to avoid confusion, one has to emphasize
that the procedure described above 
is completely  different from deriving a factorization formula 
for $B\to \pi$ form factor in terms of $B$-meson DA.
Such a formula exists \cite{BF} but it
only involves  the hard-scattering (factorizable)
part of the form factor and the DA $\phi_+$. 
In our approach, the correlator calculated in terms
of DA's is matched, via 
dispersion  relation, with the full $B\to \pi$ form factor,
or, in the absence of $\alpha_s$-corrections,
at least with the soft (nonfactorizable)
part of that form factor. 

We also presented qualitative estimates of soft-gluon 
corrections, assuming the asymptotic behaviour 
of 3-particle DA's. Our analysis indicates that 
the emission of the soft gluon contributing to the 
sum rule via 3-particle DA's, is $\sim 1/m_B$ suppressed,
and that the WW relation is violated by a small
correction. The radiative hard-gluon corrections 
to the sum rule, including renormalization effects, 
remain an important task, although
the experience with the LCSR calculation of $B\to\pi$ 
form factor tells that at finite $m_b$, the 
$O(\alpha_s)$ effects  are usually at a moderate level.

\bigskip

\noindent {\bf Acknowledgements}

\noindent We are grateful to Th.~Feldmann, M.~Beneke and V.~Braun 
for useful discussions and comments. We thank the authors of 
\cite{DFH} for informing us about their
results prior to publication. This work is supported 
by the German Ministry for Education and Research (BMBF).

\end{document}